# Spatially resolving density-dependent screening around a single charged atom in graphene

Dillon Wong[1,2]*, Fabiano Corsetti[3]*, Yang Wang[1,2]*, Victor W. Brar[1,2], Hsin-Zon Tsai[1,2], Qiong Wu[1,2], Roland K. Kawakami[4,5], Alex Zettl[1,2,6], Arash A. Mostofi[3], Johannes Lischner[3], Michael F. Crommie[1,2,6†]

[1] *Department of Physics, University of California at Berkeley, Berkeley CA, 94720, United States*
[2] *Materials Science Division, Lawrence Berkeley National Laboratory, Berkeley CA, 94720, United States*
[3] *Departments of Materials and Physics, and the Thomas Young Centre for Theory and Simulation of Materials, Imperial College London, London SW7 2AZ, United Kingdom*
[4] *Department of Physics and Astronomy, University of California, Riverside, California 92521, United States*
[5] *Department of Physics, The Ohio State University, Columbus, Ohio 43210, United States*
[6] *Kavli Energy NanoSciences Institute at the University of California, Berkeley and the Lawrence Berkeley National Laboratory, Berkeley, California 94720, United States*

\* These authors contributed equally to this work.
† Email: crommie@berkeley.edu

## Abstract

Electrons in two-dimensional graphene sheets behave as interacting chiral Dirac fermions and have unique screening properties due to their symmetry and reduced dimensionality. By using a combination of scanning tunneling spectroscopy measurements and theoretical modeling we have characterized how graphene's massless charge carriers screen individual charged calcium atoms. A back-gated graphene device configuration has allowed us to directly visualize how the screening length for this system can be tuned with carrier density. Our results provide insight into electron-impurity and electron-electron interactions in a relativistic setting with important consequences for other graphene-based electronic devices.

Understanding how screening arises from different contributions to the static dielectric function $\epsilon(q)$ is critical for unraveling material-dependent optical [1] and transport properties [2,3], as well as electron-phonon and electron-electron interactions [4,5]. Because it is two-dimensional, graphene provides a unique opportunity to study the effects of screening using spatial imaging techniques while simultaneously employing gate tunability to vary charge carrier density. The chiral relativistic nature of graphene's charge carriers [6] cause it to have a peculiar screening behavior: undoped graphene is dielectric-like while doped graphene is metal-like [7,8]. Consequently, it is possible to directly image electronic screening processes in graphene over a wide range of different screening regimes.

The screening of charged impurities is of particular importance to the performance of graphene field-effect transistors (FETs) [3,9]. Charged impurities, for example, can limit carrier mobility [2,10-12], shift the chemical potential [13], induce phase transitions [14-16], create supercritical states [17-20], and split Landau levels [21]. Although the interaction between graphene and isolated charged elements such as adsorbates [22-25] and defects [26,27] has been investigated with local probe techniques in the past, there are currently no spatially-resolved studies of the carrier-density-dependence of electronic screening of charged impurities in graphene. Here we present a systematic scanning tunneling microscopy and spectroscopy (STM/STS) study of the local screening response of gate-tunable graphene to individual charged calcium (Ca) adatoms. We find that charged impurities in graphene are screened by chiral Dirac fermions over an atypically long length scale on the order of ten nanometers. This screening length is highly dependent on carrier density and is thus tunable via gate voltage. Our spatially-resolved measurements of screening behavior in graphene are in good agreement with theoretical simulations of the electronic response of doped graphene to the presence of a screened Coulomb potential. These results demonstrate the importance of electron-electron interactions (which give rise to screening) for understanding the properties of defects in doped graphene.

We fabricated gate-tunable graphene/boron nitride (BN) devices by growing monolayer graphene via chemical vapor deposition (CVD) [28] and transferring the graphene onto BN crystals [29] exfoliated onto $SiO_2$/Si wafers. The BN flakes were used as atomically smooth substrates [30,31] with reduced charge inhomogeneity compared to $SiO_2$ [32,33]. Ca atoms were subsequently deposited onto the surface of our liquid helium cooled graphene/BN devices in an ultra-high vacuum (UHV) chamber (see Supplementary Materials and Ref. [34]). Fig. 1a depicts the graphene device used in our experimental setup. Fig. 1b shows a typical STM topographic image of graphene following this Ca deposition procedure. The Ca atoms appear as identical round protrusions on the graphene surface and are surrounded by a dark depression caused by the rearrangement of local density of states (LDOS) spectral weight above and below the Dirac point. This is a signature of the graphene screening response to the presence of charged Ca adatoms [23].

In order to determine the charge state of the Ca atoms at different doping levels we performed gate-dependent d$I$/d$V$ spectroscopy on graphene at various distances away from an isolated Ca atom (i.e., a Ca atom separated by at least 20 nm from all other Ca atoms). This data is plotted in Figs 2a-c for p-doped, nearly neutral, and n-doped graphene. Each d$I$/d$V$ curve here has been normalized by a different constant to account for the exponential dependence of the tunneling conductance on tip height [23]. All d$I$/d$V$ curves show a ~130 meV wide gap-like feature at the Fermi level caused by phonon-assisted inelastic tunneling [35,36], and the p-doped (n-doped) spectra exhibit local minima on the right (left) side of the Fermi level that reflect the graphene Dirac point. For the nearly neutral graphene spectra, the Dirac point is near the Fermi level and its location is obscured by the ~130 meV gap-like feature.

The d$I$/d$V$ curves in Figs 2a-c all display an electron-hole asymmetry in which the d$I$/d$V$ intensity at energies above the Dirac point increases as the STM tip approaches the Ca atom, while the d$I$/d$V$ intensity at energies below the Dirac point decreases as the tip approaches the Ca atom. This observation is consistent with previous theoretical predictions that the electronic LDOS of graphene increases for energies above the Dirac point as one approaches a positively charged Coulomb center while it decreases for energies below the Dirac point [18,19]. We thus conclude that the Ca atom is positively charged and stable regardless of the graphene doping level within our experimental conditions. d$I$/d$V$ spectra taken directly above individual Ca atoms confirm that there are no electronic resonances of the atom in the energy range near the Fermi level explored here, consistent with the charge stability displayed in Figs 2a-c (see Supplementary Information).

The charge stability of Ca atoms for different gating conditions allows us to image graphene's screening response to charged impurities over a wide range of doping levels. Figures 3a-c show gate-dependent d$I$/d$V$ maps near a single, positively charged Ca atom as the p-doping in graphene is progressively increased by ramping up the gate voltage (the sample bias ($V_s$) was changed at each gate voltage ($V_g$) to ensure that only electron-like states 0.15 eV above the Dirac point were tracked in all three d$I$/d$V$ maps). Figure 3a shows the d$I$/d$V$ map at the smallest gate voltage where the graphene has a p-type charge carrier density of ~3 x $10^{11}$ cm$^{-2}$. The yellow region shows the increased electron-like LDOS that occurs as graphene charge carriers rearrange themselves in response to the screened Coulomb potential of the positively charged Ca atom. Figure 3b shows the same region of graphene after raising the density of p-type charge carriers to

~$1.8 \times 10^{12}$ cm$^{-2}$. The yellow region is seen to decrease in size as the increased charge carrier density more effectively screens the Ca atom and reduces the range of its associated Coulomb potential. Figure 3c shows the same region after increasing the p-type carrier density even further to ~$3.5 \times 10^{12}$ cm$^{-2}$. Increased screening is seen to shrink the region of higher electronic LDOS around the Ca atom even further. In order to more accurately quantify these trends we measured d$I$/d$V$ line scans as a function of distance from the Ca atom. These line scans, shown in Fig. 3d, were obtained at the same energy as the d$I$/d$V$ maps of Figs. 3a-c and clarify how the graphene LDOS is modified by the screened Ca Coulomb potential for different p-type dopant levels. The characteristic decay length of the LDOS measured in the line scans is seen to decrease as the p-type graphene carrier density increases.

The results for n-doped graphene similarly show the effect on screening as charge carrier density is increased. Figures 4a-c show d$I$/d$V$ maps of the same region as Fig. 3, but for different n-doping carrier densities. Here $V_s$ was adjusted so that only hole-like states 0.08 eV below the Dirac point ($E_D$) are imaged (LDOS energies on opposite sides of $E_D$ were chosen for n- and p-doped graphene to avoid the phonon gap-like feature, thereby allowing states near $E_D$ to be characterized with greater precision). Figure 4a shows the graphene response to a single Ca atom for the smallest number of n-type charge carriers: ~$0.5 \times 10^{11}$ cm$^{-2}$. Since states below $E_D$ are imaged here the contrast is flipped compared to the images of Figs. 3a-c (we emphasize that this is not a result of the polarity of charge carriers in graphene). Figures 4b and 4c show how the n-type screening response to the Ca atom increases as carrier density is ramped up to ~$1.4 \times 10^{12}$ cm$^{-2}$. The blue region is seen to shrink as the Coulomb potential range reduces with increased screening. d$I$/d$V$ line scans were also obtained in the vicinity of the Ca atom at the same energy as the d$I$/d$V$ maps, but for different n-type carrier densities. As seen in Fig. 4d, the presence of the Ca atom strongly reduces the graphene LDOS near the atom but the LDOS returns to its unperturbed value at large distances. The length scale over which this occurs (i.e., the screening length) is seen to decrease for increased n-type carrier densities, similar to what is observed in the case of p-type charge carrier densities (Fig. 3d).

Our observation that the decay length of d$I$/d$V$ decreases with increasing carrier concentration can be qualitatively understood via Thomas-Fermi screening theory. In three-dimensional (3D) metals the static wave-vector ($q$) dependent Thomas-Fermi dielectric function is

$$\epsilon_{3D}(q) = 1 + \frac{4\pi e^2 \text{DOS}(E_F)}{q^2}, \quad (1)$$

where $\text{DOS}(E_F)$ is the density of states at the Fermi energy. However, screening in two-dimensional (2D) materials is typically weaker (resulting in stronger Coulomb interactions) because electric field lines can leave the plane of a 2D material [37]. The 2D Thomas-Fermi dielectric function is [6,7,38,39]

$$\epsilon_{2D}(q) = \epsilon_s + \frac{1}{\lambda_{TF} q}, \quad (2)$$

where $\epsilon_s$ is the effective substrate dielectric constant, and

$$\lambda_{TF} = \frac{1}{2\pi e^2 \text{DOS}(E_F)} \quad (3)$$

is the Thomas-Fermi screening length [8]. Unlike a conventional two-dimensional electron gas (2DEG) that has $\text{DOS}(E_F)$ independent of the charge carrier density $n$, graphene has a carrier-density-dependent electronic density of states and thus a carrier-density-dependent Thomas-Fermi screening length

$$\lambda_{TF} = \frac{\hbar v_F}{4e^2 \sqrt{\pi |n|}}, \quad (4)$$

where $v_F$ is the magnitude of the Fermi velocity. $\lambda_{TF}$ depends sensitively on $|n|$ and can therefore be tuned by application of a gate voltage. Increasing the magnitude of the carrier density via the gate voltage $V_g$ thus leads to a decrease of $\lambda_{TF}$, which explains the observed decrease of the decay length of d$I$/d$V$ for both p-doped (Fig. 3) and n-doped (Fig. 4) graphene.

This simple Thomas-Fermi screening picture, however, has several shortcomings. First, it does not include the effect of interband transitions between graphene's $\pi$ and $\pi^*$ bands. Second, Thomas-Fermi theory is only valid for slowly varying potentials and for energies far from the graphene Dirac point. Third, it does not directly predict the electronic LDOS, which is most closely related to the experimentally measured quantity d$I$/d$V$. Therefore, to more quantitatively and realistically explain our STM measurements, we carried out theoretical calculations for a doped graphene sheet with a single Ca adatom. We used a nearest-neighbor tight-binding model to account for graphene electronic structure and a screened Coulomb potential to describe the Ca adatom. Here the bare Coulomb potential is screened using the random phase approximation (RPA) dielectric function for the Dirac Hamiltonian [7,40]

$$\epsilon(q) = \begin{cases} \epsilon_s + \frac{2\pi e^2 \text{DOS}(E_F)}{q}, & q \leq 2k_F \\ \epsilon_s + \frac{2\pi e^2 \text{DOS}(E_F)}{q}\left[1 - \frac{1}{2}\sqrt{1-\left(\frac{2k_F}{q}\right)^2} + \frac{q}{4k_F}\cos^{-1}\frac{2k_F}{q}\right], & q > 2k_F \end{cases}, \quad (5)$$

where $k_F$ is the magnitude of the Fermi wave vector with respect to the K/K' points. The effect of changing charge carrier density in our tight-binding calculations is introduced through the dielectric function of Eq. (5). We use the following parameters in our simulation: the graphene carbon-carbon bond length $a = 0.142$ nm, $v_F = 1.1 \times 10^6$ m/s, $\epsilon_s = 2.5$, the impurity charge $Q = +0.7|e|$ (see Supplementary Information), and the height $h = 2.0$ Å of the Ca atom above the center of the graphene hexagon [41].

Figures 2d-f show the results of our simulated d$I$/d$V$ point spectra for p-doped, nearly neutral, and n-doped graphene (each colored curve corresponds to a different distance from the Ca atom). Quasiparticle lifetime effects and inelastic tunneling processes have been included (see Ref. [36] for details on this procedure; the Supplementary Information shows theoretical curves without lifetime and inelastic tunneling effects). In agreement with the experimental data (Figs 2a-c), the computed spectra exhibit a significant electron-hole asymmetry when the tip is brought closer to the adatom; the simulated LDOS increases above the Dirac point and decreases below the Dirac point for closer distances.

An intuitive picture for understanding these findings is that the LDOS of graphene in the presence of the charged impurity is described by the LDOS of unperturbed graphene, but shifted towards lower energies by the local value of the screened Coulomb potential. This explains the reduction of dI/dV below the Dirac point and its increase above the Dirac point. We find that a shifted LDOS is in good agreement with our calculations for energies sufficiently far from the Dirac point (see Supplementary Information). In the vicinity of the Dirac point, however, this intuitive picture breaks down. In particular, the Dirac point itself does not shift in energy – a consequence of the linear dispersion of the graphene Dirac bands [18].

To model our experimental d$I$/d$V$ maps and better visualize the spatial dependence of the screening behavior we calculated the theoretical tunneling conductance as a function of distance away from a Ca adatom at fixed energy. Figures 3e and 4e show simulated d$I$/d$V$ versus distance for p-doped and n-doped graphene, respectively. The energies and charge carrier densities $n$ were chosen such that Fig. 3e directly corresponds to Fig. 3d, and Fig. 4e to Fig. 4d. In agreement with the experimental results shown in Figs 3d and 4d, the theoretical spatial profile of the tunnel conductance decays more rapidly for higher doping levels (for both p-doped and n-doped graphene), directly reflecting the reduced range of the impurity potential caused by a reduced screening length. We also carried out large-scale first-principles calculations of the calcium-graphene system within a density functional theory (DFT) framework as implemented in the ONETEP code [42,43], which confirm the trends obtained from the tight-binding model (see Supplementary Information).

These results directly confirm that the RPA model accurately describes screening in graphene. RPA screening has already played an essential ingredient in early theoretical models of bipolar electron transport in graphene, as it explains the V-shaped conductivity as a function of gate voltage [44-46]. Screening of charged impurities causes long-range impurity scattering to dominate graphene's transport properties at low carrier concentration and short-range impurity scattering to dominate at high carrier concentration [2]. Our data for the simplest possible charged impurity system – a single, isolated impurity on graphene – allows us to directly visualize this phenomenon and confirm these assumptions.

In conclusion, we have explored how relativistic charge carriers in graphene screen a single Coulomb potential for different carrier densities. Direct STM/STS measurements of the local electronic structure of gate-tunable graphene in the presence of isolated Ca adatoms have allowed us to directly observe how the screening length of graphene decreases with increasing charge carrier density. Unlike conventional 2DEGs, where the Thomas-Fermi screening length $\lambda_{TF}$ is independent of carrier density [47], graphene's Dirac-like band structure leads to $\lambda_{TF} \propto 1/\sqrt{|n|}$. This experimental trend is confirmed by a tight-binding model of graphene incorporating a screened Coulomb potential. The fundamental behavior described here (as well as visualization techniques) can be generalized to other electrostatic potentials, such as graphene pn junctions [48-51], quantum dots [52,53], and superlattices [54-60] where the potential landscape felt by graphene charge carriers is altered by density-dependent screening effects.

Data underlying this article can be accessed on figshare at https://dx.doi.org/10.6084/m9.figshare.3824451, and used under the Creative Commons Attribution license.  This work was supported by the sp$^2$-bonded materials program (KC2207) (STM measurement and instrumentation development) funded by the Director, Office of Science, Office of Basic Energy Sciences, Materials Sciences and Engineering Division, of the US Department of Energy under Contract No. DE-AC02-05CH11231.  For the graphene characterization we used the Molecular Foundry at LBNL, which is funded by the Director, Office of Science, Office of Basic Energy Sciences, Scientific User Facilities Division, of the US Department of Energy under Contract No. DE-AC02-05CH11231.  Support was also provided by National Science Foundation award DMR-1206512 (device fabrication, image analysis).  F.C. and A.A.M. were supported by the EPSRC under Grant No. EP/J015059/1 (density functional theory calculations).  J.L. acknowledges support from EPSRC under Grant No. EP/N005244/1 (tight-binding calculations).  F.C., A.A.M., and J.L. acknowledge support from the Thomas Young Centre under grant no. TYC-101 and the Imperial College London High Performance Computing Service (numerical algorithm development).  This work used the ARCHER UK National Supercomputing Service via J.L.'s membership of the UK's HEC Materials Chemistry Consortium.  D.W. was supported by the U.S. Department of Defense through the National Defense Science & Engineering Graduate Fellowship (NDSEG) Program, 32 CFR 168a.  We thank L.S. Levitov, A.V. Shytov, and V.M. Pereira for helpful discussions.

**FIGURE 1.** Calcium adatoms on graphene. (a) Schematic of experimental setup. Calcium atoms are deposited onto a graphene/BN/SiO$_2$/Si FET device. A voltage $V_g$ is applied to Si to tune the charge carrier density in graphene, and a voltage $-V_s$ is applied to the STM tip. (b) STM topographic image of Ca atoms adsorbed onto a graphene/BN surface. Tunneling parameters: $V_s$ = -0.45 V, $I$ = 2 pA.

**FIGURE 2.** STS measurements near an individual calcium atom and theoretical simulations. (a) Normalized d$I$/d$V$ point spectra measured at different distances from a single Ca atom on p-doped graphene. (b) Same as (a) for a Ca atom on nearly neutral graphene. (c) Same as (a) for a Ca atom on n-doped graphene. These d$I$/d$V$ spectra show that a Ca atom on graphene remains positively charged as graphene's charge carrier density is tuned via a back-gate voltage $V_g$. Initial tunneling parameters: (a) $V_s$ = 0.6 V, $I$ = 60 pA, $V_g$ = -60 V; (b) $V_s$ = 0.6 V, $I$ = 60 pA, $V_g$ = -30 V; (c) $V_s$ = 0.6 V, $I$ = 60 pA, $V_g$ = 30 V. (d) Tight-binding simulation of p-doped graphene d$I$/d$V$ point spectra at different distances from a screened Coulomb potential as described in text. (e) Same as (d) for a screened Coulomb potential in nearly neutral graphene. (f) Same as (d) for a screened Coulomb potential in n-doped graphene. The Dirac points in (a), (c), (d), and (f) are indicated by black arrows.

**FIGURE 3.** Gate-dependent d$I$/d$V$ maps for p-doped graphene. (a) d$I$/d$V$ map of electronic states 0.15 eV above the Dirac point in the vicinity of a single Ca atom (represented by red disk) on p-doped graphene: $V_s$ = 0.28 V, $I$ = 28 pA, $V_g$ = 0 (the Ca atom was not directly scanned in order to minimize the risk of picking the atom up with the STM tip). (b) Same as (a) but with $V_s$ = 0.38 V, $I$ = 38 pA, $V_g$ = -30V. (c) Same as (a) but with $V_s$ = 0.45 V, $I$ = 45 pA, $V_g$ = -60V. (d)

Radially averaged d$I$/d$V$ linecut of electronic states 0.15 eV above the Dirac point as a function of distance from a single Ca atom on p-doped graphene (the data is normalized to account for tip height variations caused by the STM feedback loop as it maintained constant current [23]). Curves are vertically offset for clarity, with the magnitude of p-doping increasing from top curve to bottom curve. The value of d$I$/d$V$ far from the Ca atom is set to 1. (e) Simulated d$I$/d$V$ linecuts of electronic states 0.15 eV above the Dirac point as a function of distance from an RPA-screened Coulomb potential on p-doped graphene. Charge carrier density values for each line cut were chosen to correspond to the gate voltages in (d). Graphene here is modeled using tight-binding theory. The value of the simulated d$I$/d$V$ far from the screened Coulomb potential is normalized to 1.

**FIGURE 4.** Gate-dependent d$I$/d$V$ maps for n-doped graphene. (a) d$I$/d$V$ map of electronic states 0.08 eV below the Dirac point in the vicinity of a single Ca atom (represented by red disk) on n-doped graphene: $V_s$ = -0.16 V, $I$ = 17 pA, $V_g$ = 5V (the Ca atom was not directly scanned in order to minimize the risk of picking the atom up with the STM tip). (b) Same as (a) but with $V_s$ = -0.22 V, $I$ = 20 pA, $V_g$ = 20 V. (c) Same as (a) but with $V_s$ = -0.28 V, $I$ = 28 pA, $V_g$ = 40 V). (d) Radially averaged d$I$/d$V$ linecuts of electronic states 0.08 eV below the Dirac point as a function of distance from a single Ca atom on n-doped graphene (the data is normalized to account for tip height variations caused by the STM feedback loop as it maintained constant current [23]). Curves are vertically offset for clarity, with the magnitude of n-doping increasing from top curve to bottom curve. The value of d$I$/d$V$ far from the Ca atom is set to 1. (e) Simulated d$I$/d$V$ linecuts of electronic states 0.08 eV below the Dirac point as a function of distance from an RPA-screened Coulomb potential on n-doped graphene. Charge carrier density

values for each linecut were chosen to correspond to the gate voltages in (d). Graphene here is modeled using tight-binding theory. The value of the simulated d$I$/d$V$ far from the screened Coulomb potential is normalized to 1.

# FIGURE 1

**a**

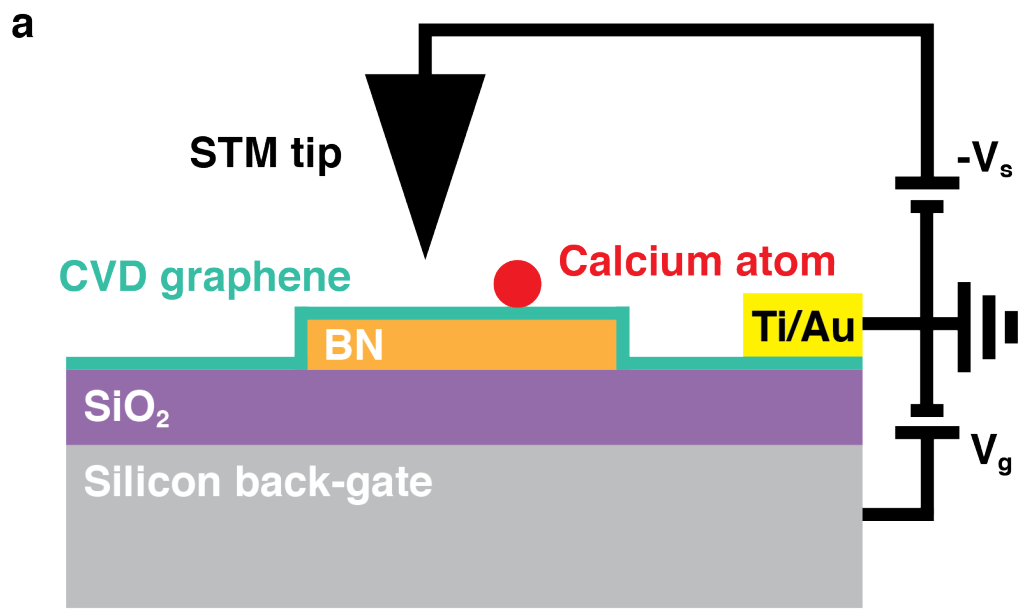

**b**

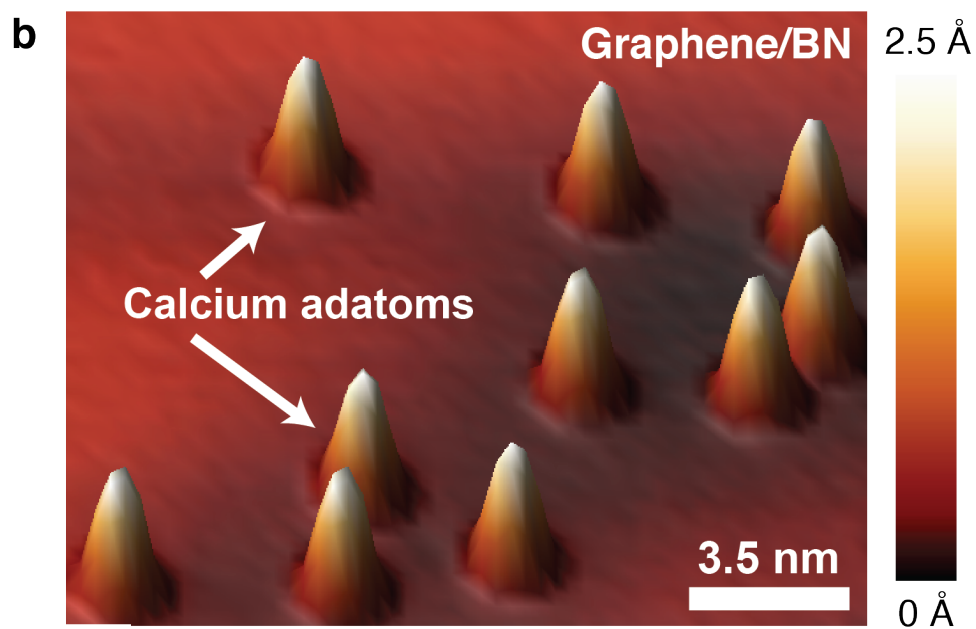

# FIGURE 2

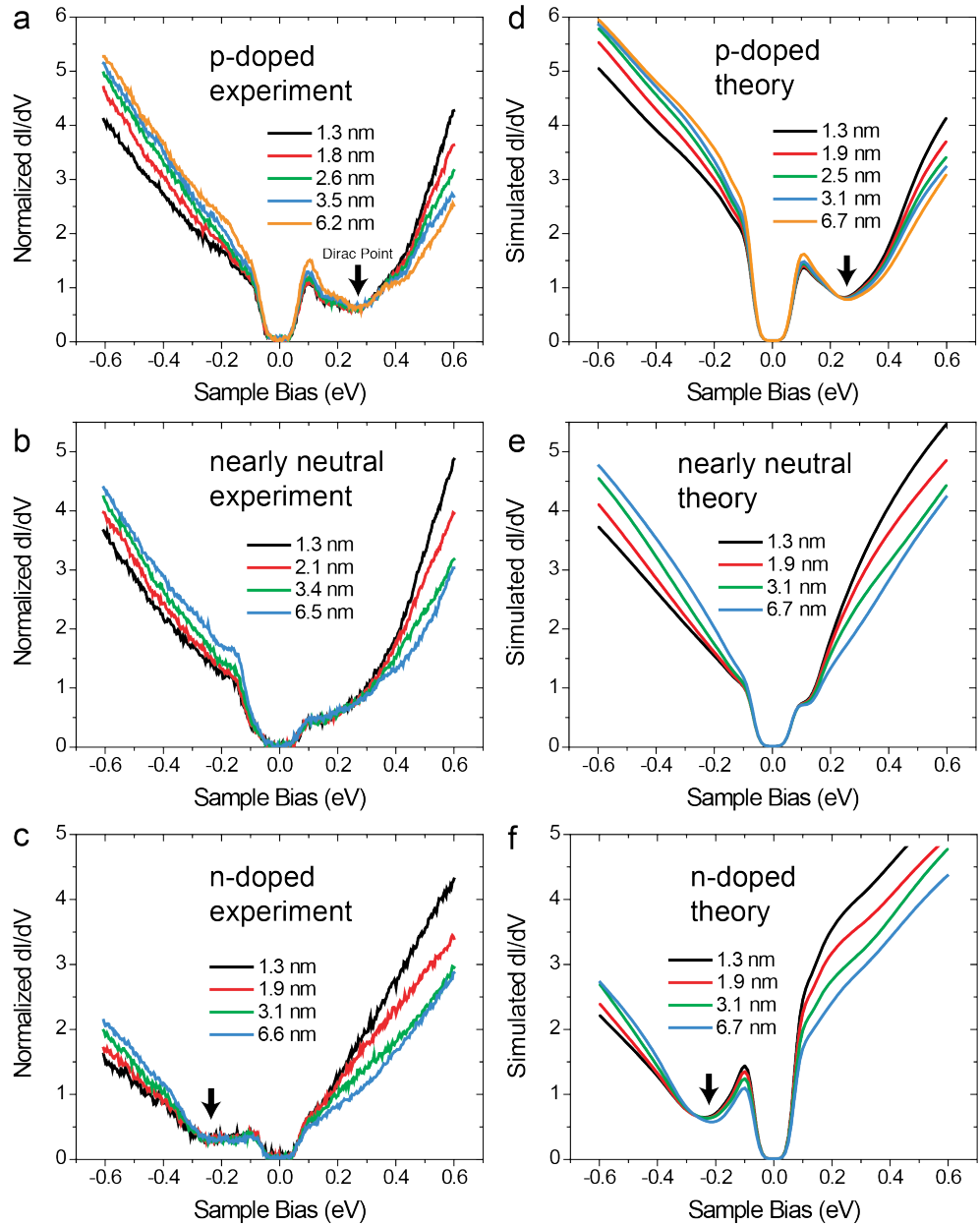

# FIGURE 3

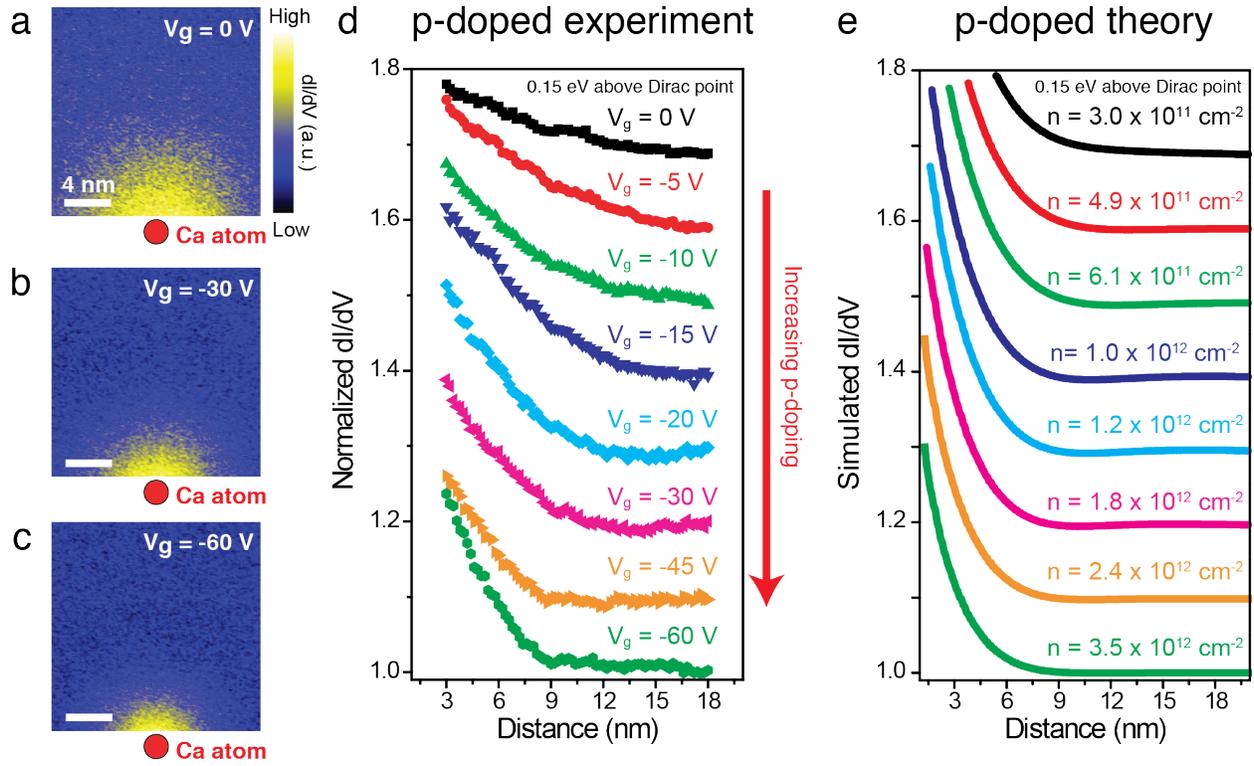

# FIGURE 4

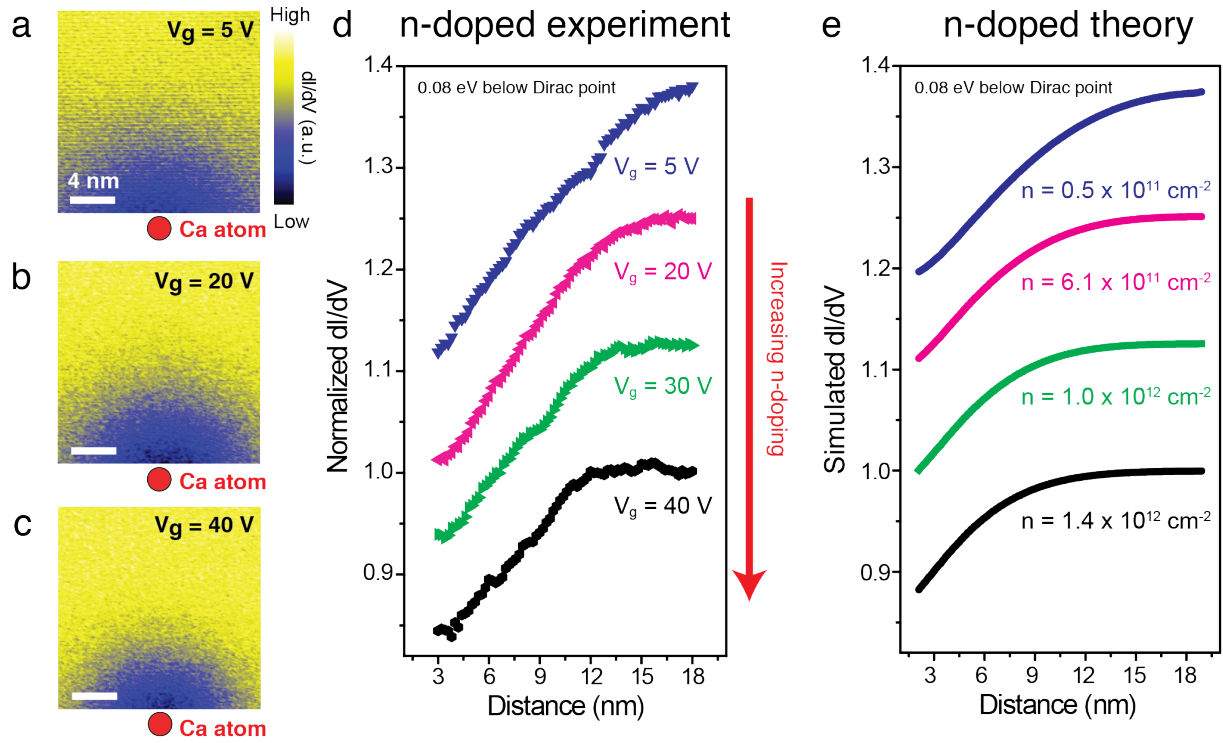